\documentclass[smallextended]{svjour3}

\usepackage{graphicx}

\usepackage{mathptmx}

\journalname{Gen. Rel. Grav.}

\newcommand{\G}{\mathcal{G}}

\date{Received: \today}

\begin{document}

\title{Cosmological scaling solutions in generalised Gauss-Bonnet
gravity theories}


\author{Kotub Uddin \and James E. Lidsey \and Reza Tavakol}


\institute{Astronomy Unit, \\
School of Mathematical Sciences, \\
Queen Mary University of London,\\
London E1 4NS, United Kingdom \\
\email{J.E.Lidsey@qmul.ac.uk} \\
\email{k.uddin@qmul.ac.uk} \\
\email{r.tavakol@qmul.ac.uk} \\
}

\maketitle

\begin{abstract}
\large
The conditions for the existence and 
stability of cosmological power-law scaling solutions 
are established when the Einstein-Hilbert action is modified 
by the inclusion of a 
function of the Gauss-Bonnet curvature invariant. The general form of the 
action that leads to such solutions is determined for the case 
where the universe is sourced by a barotropic perfect fluid. 
It is shown by employing an equivalence between the Gauss-Bonnet 
action and a scalar-tensor theory of gravity that the cosmological
field equations can be written as a plane autonomous system. 
It is found that stable scaling solutions exist when the parameters of the 
model take appropriate values. \\

\end{abstract}

\section{Introduction}
\large
In recent years, there has been considerable interest in the possibility
that Einstein's theory of general relativity may become modified
in high-curvature regimes and over large distance scales. This
possibility has been motivated by a wealth of high redshift observations,
which indicate that the universe is presently undergoing a phase of
accelerated expansion \cite{SnIa1,SnIa2,SnIa3,SnIa4,SnIa5,SnIa6,Fay06,CMB1,CMB3,CMB5,LSS1,LSS2,LSS3,LSS4,LSS5}.
(For a review, see, e.g. \cite{Review}). In many such studies,
the Einstein-Hilbert action is modified by the
introduction of terms involving
higher-order curvature invariants. An important quadratic
combination of such invariants, which is motivated
by string theory, is given by the Gauss-Bonnet (GB) invariant
\begin{equation}
\label{GBI}
\G \equiv R^2 -4R^{\mu\nu}R_{\mu\nu}+R^{\mu\nu\rho\tau}R_{\mu\nu\rho\tau}.
\end{equation}
In four dimensions, the GB term is a topological invariant and
introducing a term proportional to $\G$ into the Einstein-Hilbert
action does not modify the dynamics. Recently, however, the
cosmology of models based on a class of generalised
theories with an action of the form
\begin{equation}
\label{MGBact}
S=\int d^{4}x \sqrt{-g} \left(\frac{R}{2}+f(\G)\right)
+S_m 
\end{equation}
has been considered, where $f (\G )$ is a differentiable function
of $\G$ and $S_m$ represents the matter action
\cite{Nor-Od-O05,Nojiri06,Cognola,Cognola06,fG,Davis07,DeFelice07,Cognola07,Brevik,Bamba,DeFelice,Bazeia,Ali,Boehmer,Zhou}.

The purpose of the present paper is to investigate the
existence and stability of cosmological power-law 
scaling solutions derived from
theories of the type (\ref{MGBact}) in the presence of a perfect fluid matter
source. Scaling (attractor) solutions play an important role in cosmology,
since they enable the asymptotic behaviour and stability
of a particular cosmological background to be determined. Moreover,
they provide a framework for establishing the behaviour of more
general cosmological solutions \cite{Copeland98,Liddle99,Rubano01,Copeland04,Tsujikawa04,Steinhardt}.

The structure of the paper is as follows.
We begin in Section 2 by summarizing the derivation of the
cosmological field equations by employing an equivalence between the
action (\ref{MGBact}) and a corresponding action involving a self-interacting
scalar field that is non-minimally coupled to gravity.
We focus on the spatially flat and isotropic
Friedmann-Lemaitre-Robertson-Walker (FLRW) universe and
proceed in Section 3 to
identify the most general form for the function $f(\G )$ that results 
in power-law (scaling) solutions when
the matter source is a barotropic fluid with a constant equation
of state parameter. Specifically, we find that
scaling solutions may arise when $f =\pm 2 \sqrt{\alpha\G}$,
where $\alpha$ is an arbitrary constant. We then
show that for this form of the action, the field equations can be
expressed as a plane autonomous system.
This allows us to employ dynamical systems theory to investigate
the stability of the vacuum and non-vacuum solutions and
this is done in Sections 4 and 5, respectively.
We find that scaling solutions, corresponding either to a stable node or
a stable spiral node, can arise when the equation of state of the fluid
and the parameter, $\alpha$, satisfy appropriate conditions.
We conclude with a discussion in Section 6.  Units are chosen such
that $8\pi G = c =1$.

\section{Cosmological Field Equations}

Action (\ref{MGBact}) may be expressed in an alternative form
by introducing two auxiliary scalar fields $\chi$ and $\zeta$
such that \cite{Nor-Od-O05,SotiriouThesis,Wands,Chiba}
\begin{equation}
\label{auxact}
S=\int d^{4}x \sqrt{-g} \left(\frac{R}{2}+\zeta(\G-\chi)
+f(\chi)\right)+S_m.
\end{equation}
Varying Eq. (\ref{auxact}) with respect to $\zeta$ yields the constraint
$\chi=\G$, thereby
reproducing action (\ref{MGBact}). On the other hand, varying action
(\ref{auxact})
with respect to $\chi$ implies that
$\zeta=F(\chi)$, where $F(\chi)\equiv \partial f(\chi)/\partial \chi$, and
substituting this condition back into Eq. (\ref{auxact}) leads to
\begin{equation}
\label{auxact1}
S=\int d^{4}x \sqrt{-g} \left(\frac{R}{2}
+F(\chi)(\G-\chi)+f(\chi)\right)+S_m.
\end{equation}
It follows, therefore, that the action (\ref{MGBact})
is equivalent to the action \cite{Nor-Od-O05,SotiriouThesis}
\begin{equation}
\label{GBequivact}
S=\int d^{4}x \sqrt{-g} \left(\frac{R}{2}-V(\phi)
-h(\phi)\G\right) +S_m , 
\end{equation}
where the scalar field, $\phi$, is defined implicitly
by
\begin{equation}
\label{defh}
h(\phi ) \equiv -F(\G )
\end{equation}
for some function $h(\phi)$
and has an effective self-interaction potential
\begin{equation}
\label{potn}
V(\phi) \equiv \G F(\G)-f(\G),
\end{equation}
where $F \equiv \partial f /\partial \G$. Eq. (\ref{GBequivact})
may be interpreted as an effective `scalar-tensor' theory, where the scalar
field has a vanishing kinetic term.

To study cosmological models based on action (\ref{MGBact}),
one may proceed directly by varying the action to derive
the field equations or, indirectly, by
varying the equivalent action (\ref{GBequivact}). We employ the
latter approach in the present work in view of its potential simplicity.
The field equations in this case take the form
\begin{equation}
\label{FE1}
R^{\mu\nu} -\frac{1}{2} R g^{\mu\nu} =
 T_{m}^{\mu\nu}+T_{\G}^{\mu\nu}
\end{equation}
where $T_{m}^{\mu\nu}$ is the energy-momentum tensor of the matter
fields and
$T_{\G}^{\mu\nu}$ denotes the effective energy-momentum tensor
resulting from the scalar field, $\phi$, and the GB term.
Since the GB term is a topological invariant in four dimensions,
the standard field equations of GR are recovered when
$h(\phi)={\rm constant}$. Consequently, only terms involving
derivatives of $h(\phi)$  arise in the energy-momentum tensor, which
is given by \cite{Sotiriou06}
\begin{eqnarray}
\label{EM}
T_{\G}^{\mu\nu}
=&-&g^{\mu\nu}V(\phi)
-2[\nabla^{\mu}\nabla^{\nu}h(\phi)]R
+2g^{\mu\nu}[\nabla^2h(\phi)]R
+4[\nabla_{\rho}\nabla^{\mu}h(\phi)]R^{\nu\rho} \nonumber
\\
&+&4[\nabla_{\rho}\nabla^{\nu}h(\phi)]R^{\mu\rho}
-4[\nabla^{2}h(\phi)]R^{\mu\nu}
-4g^{\mu\nu}[\nabla_{\rho}\nabla_{\tau}h(\phi)]R^{\rho\tau} \nonumber \\
&+&4[\nabla_{\rho}\nabla_{\tau}h(\phi)]R^{\mu\rho\nu\tau}.
\end{eqnarray}
Finally, the equation of motion for the scalar field takes the form
\begin{equation}
\label{FE2}
V_{,\phi}(\phi)+h_{,\phi}(\phi)\G=0,
\end{equation}
where a comma denotes differentiation with respect to $\phi$.

Our aim is to study the dynamics of the isotropic and spatially flat
FLRW universe sourced
by a perfect barotropic fluid with an equation of state parameter,
$w_m = p_m/\rho_m$, where $p_m$ and $\rho_m$ denote
the pressure and energy density of the fluid, respectively.
For this spacetime, the GB invariant is given by
$\G=24H^2(\dot{H}+H^2)$, where
$H\equiv \dot{a}/{a}$ defines the Hubble parameter,
$a$ represents the scale factor of the universe and
a dot denotes differentiation with respect
to cosmic time. The Friedmann and Raychaudhuri
equations derived from Eqs. (\ref{FE1})-(\ref{EM}) for this background
are then given by \cite{Cognola,Neupane06}
\begin{eqnarray}
\label{Friedmann}
3H^2&=&V(\phi) +24H^3\dot{h} +\rho_m,
\\
\label{Raychoudhuri}
\left(2\frac{\dot{H}}{H^2}+3\right)H^2&=&V(\phi)
+ 8H^2\ddot{h}+
16H^3\dot{h}\left(1+\frac{\dot{H}}{H^2}\right)-p_m ,
\end{eqnarray}
respectively, and the scalar field equation (\ref{FE2}) reduces to
\begin{equation}
\label{K-G}
V_{,\phi}+24h_{,\phi}H^2(\dot{H}+H^2)=0.
\end{equation}

It proves convenient to interpret the GB gravitational
terms on the right-hand side of the Friedmann
equation (\ref{Friedmann}) as an effective energy density, such that
$\rho_{\G} \equiv T_{\G} +V(\phi )$, where $T_{\G} \equiv 24\dot{h}H^3$
plays the role of a kinetic energy. It is then natural to
introduce the dimensionless variables
\begin{equation}
\label{var}
y_1\equiv \frac{V(\phi)}{3H^2},~~~y_{2}\equiv8H \dot{h},
\end{equation}
and the fractional energy densities
\begin{eqnarray}
\label{var2}
\Omega_{m} &\equiv& \frac{\rho_m}{3H^2}=1-y_1-y_2,\\
\label{var2a}
\Omega_{\G} &\equiv& y_1+y_2.
\end{eqnarray}
The background field
equations~(\ref{Friedmann})-(\ref{K-G})
can then be expressed in terms of these variables such that
\begin{eqnarray}
\label{autn}
\frac{dy_1}{dN}&=& 2\epsilon y_1-(1-\epsilon)y_2, \\
\label{autna}
\frac{dy_2}{dN}&=& -2\epsilon+3(1-y_1)-(2-\epsilon)y_2+3w_m\Omega_m,
\end{eqnarray}
where  $\epsilon \equiv -\dot{H}/H^2$ and $N \equiv \ln a$.

\section{Cosmological Scaling Solutions}

We wish to identify the class of
GB theories that admit scaling solutions such that each of the terms in the
Friedmann equation (\ref{Friedmann}) scales at the same rate,
$H^2 \propto \rho_m \propto V(\phi) \propto T_{\G}$ \cite{Tsujikawa06}.
These conditions result
in a power-law solution to Eqs. (\ref{Friedmann})-(\ref{K-G})
of the form $a \propto t^{1/\epsilon}$, where
$\epsilon = {\rm constant}$.
For such a scaling solution, it follows from Eq. (\ref{K-G}) that
\begin{equation}
\label{prop}
V_{,\phi} = -\frac{1}{\alpha} V^2h_{,\phi}
\end{equation}
when $\epsilon \ne  1$, where $\alpha$ is a finite constant.
Integrating Eq. (\ref{prop}) then implies that
\begin{eqnarray}
\label{V-h}
h = \frac{\alpha}{V} +\beta,
\end{eqnarray}
where $\beta$ is an
arbitrary integration constant.

Relating the functions $V(\phi)$ and $h(\phi)$ in this way is equivalent
to specifying the form of the GB function, $f(\G)$, via the definition
given in Eq. (\ref{potn}). Indeed,
substituting Eq. (\ref{V-h}) into Eq. (\ref{potn})
results in the first-order, non-linear differential equation
\begin{equation}
\label{clairaut}
\left( \G \frac{df}{d\G} -f \right) \left( \frac{df}{d\G} + \beta \right)=-\alpha.
\end{equation}
Eq. (\ref{clairaut}) is an example of Clairaut's equation \cite{Ince} and may
be solved in full generality by differentiating with respect to $\G$:
\begin{equation}
\label{diff}
\frac{d^2f}{d \G^2} \left[ \left( \frac{df}{d\G} +\beta \right)^2
- \frac{\alpha}{\G} \right] =0.
\end{equation}
Eq. (\ref{diff}) is trivially solved by $f(\G ) = \alpha_0 +\alpha_1\G$,
where $\alpha_i$ are constants.
However, this simply corresponds to the introduction
of a cosmological constant in the action (\ref{MGBact}) and is not physically
interesting to the present discussion. (Recall that a contribution
of the form $f\propto \G$ is also uninteresting since the GB
term is a topological invariant).
On the other hand, a singular solution to
Eq. (\ref{clairaut}) with no arbitrary
constants can be found by setting the square
bracketed term in Eq. (\ref{diff}) to zero and substituting the result into
Eq. (\ref{clairaut}). We find that
\begin{equation}
\label{model}
f(\G)= \pm2 \sqrt{\alpha \G} ,
\end{equation}
where we have specified $\beta =0$ without loss of generality. 
Moreover, requiring the action (\ref{MGBact}) to be real
implies that $\alpha\G >0$.

Eqs. (20) and (23) represent the necessary and sufficient 
conditions for the existence of power-law scaling solutions, where 
$\epsilon= {\rm constant}$. More general solutions to the field 
equations, where $\epsilon$ is time-dependent, 
exist for this model.
If the cosmological behaviour of the model (\ref{model}) is to be determined, 
the coupled differential equations (\ref{autn})-(\ref{autna})
must close. This implies that the parameter $\epsilon$ must be
expressible as a function of $y_1$ and $y_2$ only. When Eq. (\ref{V-h})
is satisfied, we find that
\begin{eqnarray}
\label{epsScal}
\epsilon=1-\frac{3}{8\alpha}y_1^2.
\end{eqnarray}
Hence, substituting Eq. (\ref{epsScal}) into
Eqs. (\ref{autn})-(\ref{autna}) yields the plane autonomous system:
\begin{eqnarray}
\label{autn1}
\frac{dy_1}{dN}&=& 2y_1-\frac{3}{4\alpha} y_1^3-\frac{3}{8\alpha} y_1^2y_2,
\\
\label{autn1a}
\frac{dy_2}{dN}&=& 2(y_2-1)-\frac{3}{8\alpha} y_1^2y_2+\frac{3}{4\alpha}
y_1^2+3(1+w_m)(1-y_1-y_2).
\end{eqnarray}

Before concluding this section, it should be remarked that the
equivalence between actions (\ref{MGBact}) and (\ref{GBequivact})
does not apply for the special case $\epsilon =1$ $(y_1 =0)$,
corresponding to
the coasting solution, $a \propto t$. In this case,
integration of Eq. (\ref{K-G}) would
yield $V(\phi) = V_0 = {\rm constant}$ and the solution to
Eq. (\ref{potn}) would then be given by
$f( \G ) = -V_0 +\gamma \G$ for some constant $\gamma$.
This disparity can be traced to the singular nature of the coasting solution
for the model (\ref{model}). Specifically, the Friedmann equation derived
directly from action (\ref{MGBact}) for this  model is given by
\begin{equation}
\label{FRD2}
3H^2 = \mp \sqrt{6\alpha} \frac{H^2(2H^3-\ddot{H})}{(\dot{H}+H^2)^{3/2}}
+ \rho_m
\end{equation}
and the term originating from the GB contribution is ill-defined
when $\epsilon =1$ $(y_1 =0)$. Consequently,
we do not consider this solution in the phase plane analyses of the
following sections.

\section{Vacuum solutions}

In this Section, we consider vacuum solutions where
$\Omega_m =0$ and $y_1=1-y_2$. The
pair of equations (\ref{autn1})-(\ref{autn1a})
then reduces to the one-dimensional system
\begin{eqnarray}
\label{autnvac}
\frac{dy_1}{dN}= y_1 \left(2-\frac{3}{8\alpha} y_1-\frac{3}{8\alpha} y_1^2\right).
\end{eqnarray}

\begin{figure}[t]
\includegraphics[width=2.1in,height=2in]{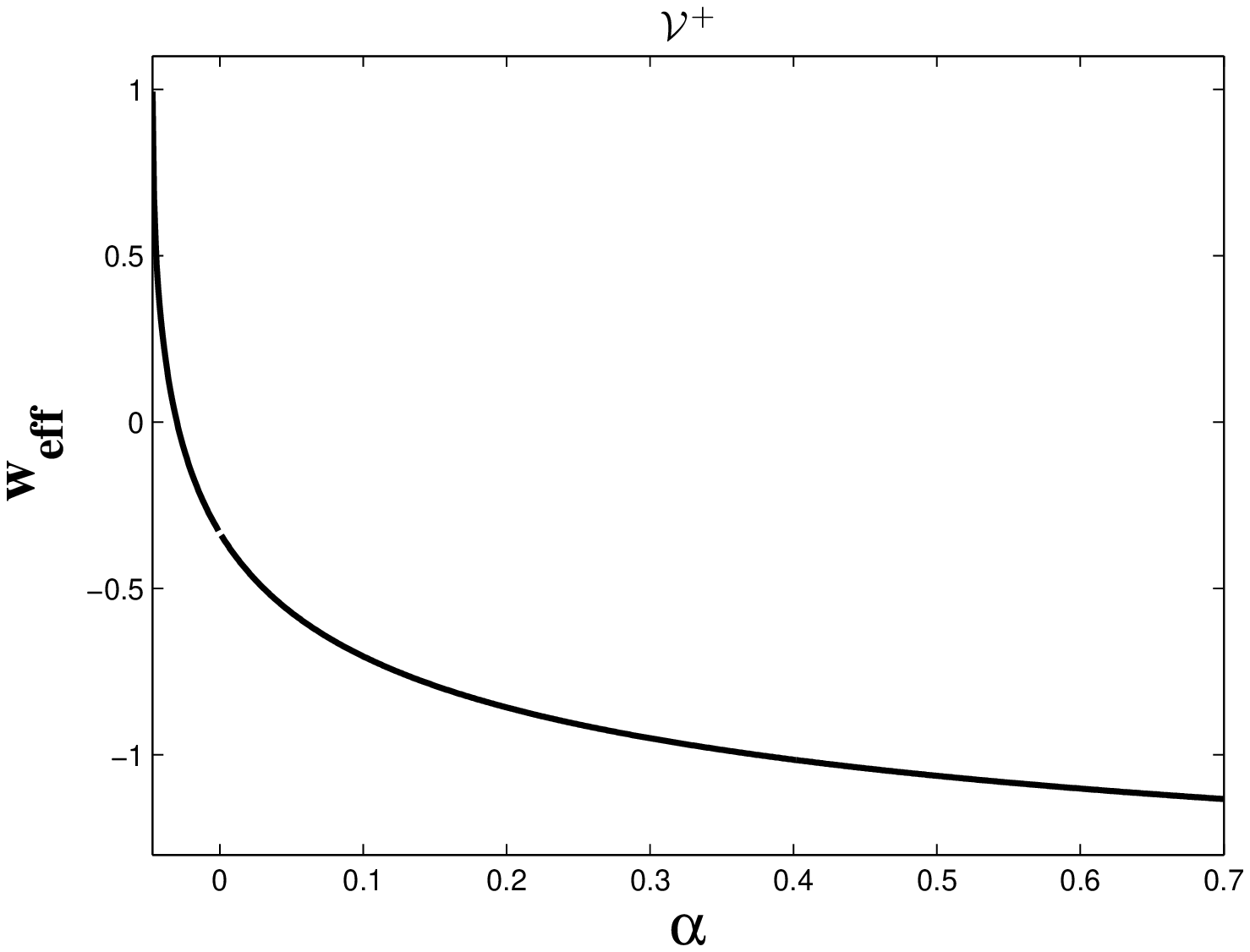}\includegraphics[width=2.1in,height=2in]{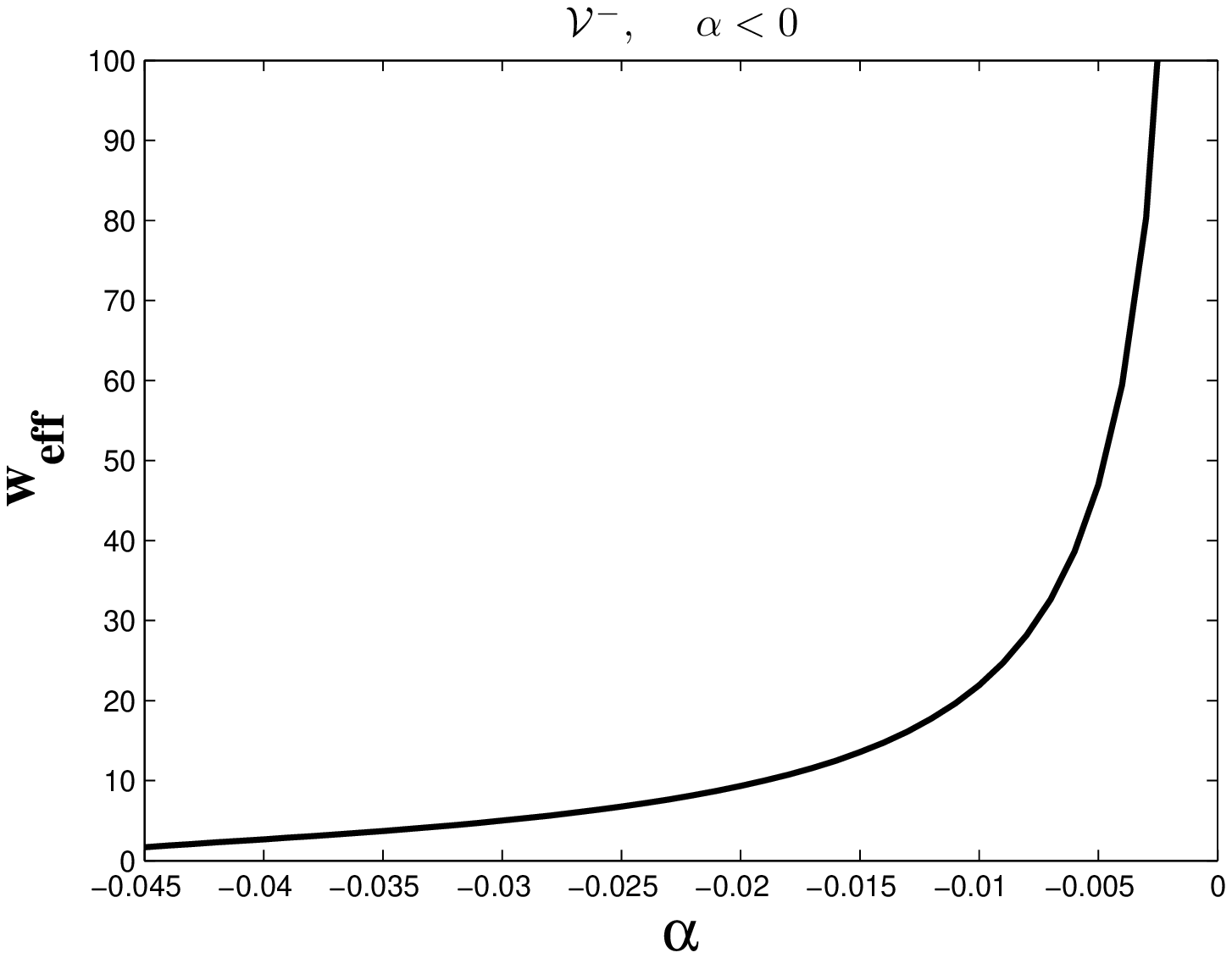}\includegraphics[
width=2.1in,height=2in]{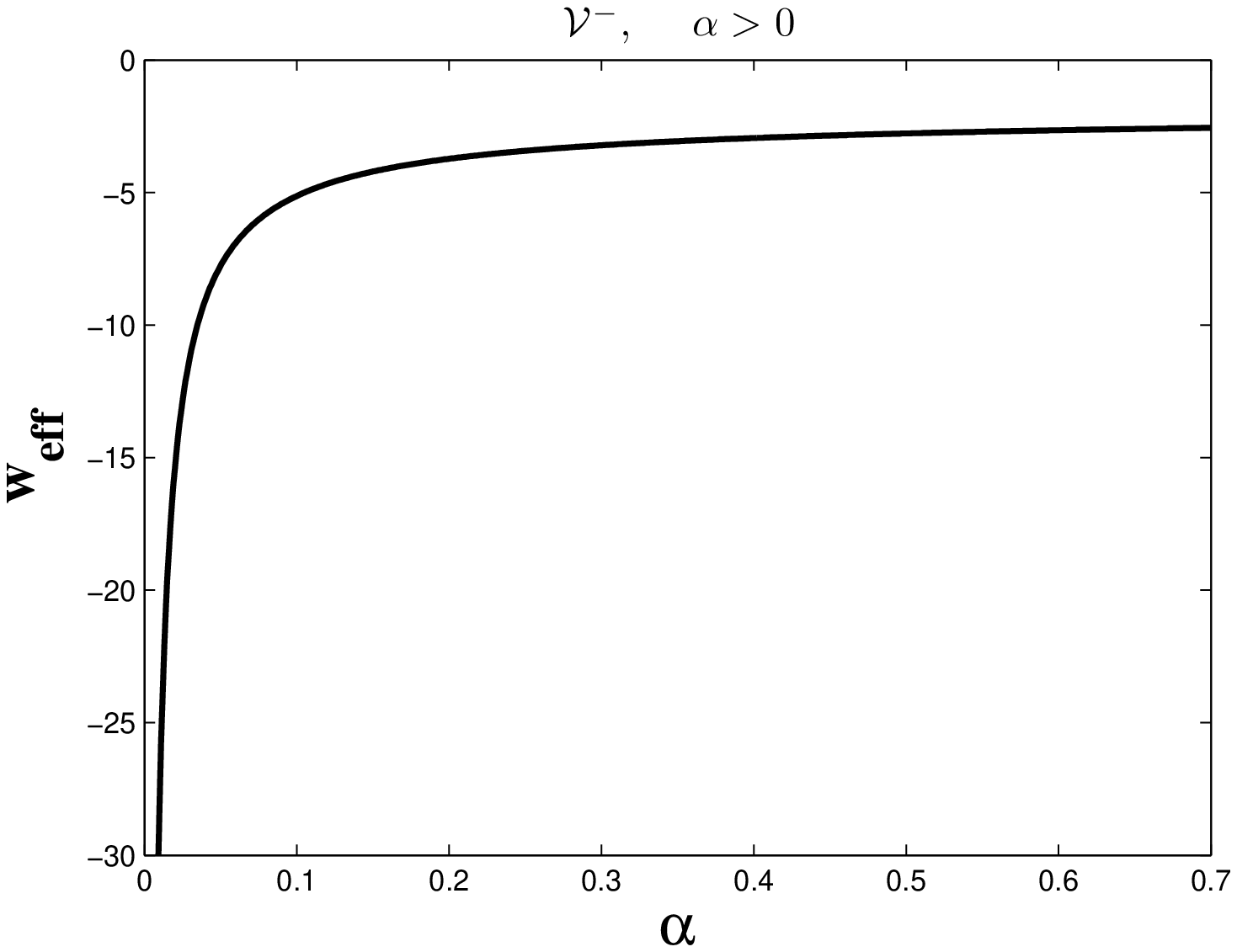}
\caption{\label{fig1}
Illustrating the effective equation of state $w_{eff}$
for the vacuum solutions $\mathcal{V}^+$ and
$\mathcal{V}^-$. Requiring that the fixed points are real
yields the condition $\alpha\geq -9/192$.
The left-hand panel corresponds to $\mathcal{V}^+$,
which shows that $\mathcal{V}^+$
is an accelerating solution for $\alpha>0$ and
corresponds to the de Sitter solution if $\alpha=3/8$.
The middle panel corresponds to $\mathcal{V}^-$
when $\alpha<0$ and in this regime $w_{eff}\geq 1$.
The right-hand panel corresponds to $\mathcal{V}^-$
when $\alpha>0$ and in this regime $w_{eff}<-1$.
}
\end{figure}

There exist two power-law solutions when $y_1 \ne  0$:
\begin{equation}
\label{vacv1}
y_1 = -\frac{1}{2}\pm\frac{1}{6}\sqrt{9+192\alpha},
\end{equation}
which we denote as $\mathcal{V}^\pm$, respectively.
The reality of the fixed points requires that $\alpha \ge -9/192$.
The power of the expansion can be expressed
in terms of the effective equation of state parameter
\begin{equation}
\label{weff}
w_{eff}\equiv -1+\frac{2}{3}\epsilon
\end{equation}
such that $a(t) \propto t^{2/[3(1+w_{eff})]}$.
It is determined by the value of the GB coupling parameter, $\alpha$, and 
substituting Eqs. (\ref{epsScal})
and (\ref{vacv1}) into Eq. (\ref{weff}) implies that
\begin{equation}
\label{weffc}
w_{eff}=\frac{1}{24\alpha} \left[ -40\alpha-3 \pm \sqrt{9+192\alpha} \right] ,
\end{equation}
where the $+/-$ corresponds to the points ${\cal{V}}^{\pm}$,
respectively. This dependency of the effective equation of state on
the GB parameter is illustrated in Fig. \ref{fig1}.
The solution $\mathcal{V}^+$
corresponds to an inflationary cosmology
when $\alpha>0$ and the exponential, de Sitter solution
arises when $\alpha=3/8$. The solution $\mathcal{V}^-$
is in a super-inflationary regime
($w_{eff}<-1$) for $\alpha>0$. When $\alpha <0$, the effective equation
of state corresponds to that of an ultra-stiff fluid
($w_{eff}\geq 1$).
Our results are in line with the recent conclusions of Ref.~\cite{Ishak},
where a study of the late-time cosmology based on the
model $f(\G) \propto -\G^n$ was made with the
field equations derived directly from action (\ref{MGBact}).

The eigenvalues associated with the equilibrium points
${\cal{V}}^{\pm}$ are given by
\begin{equation}
\label{evacv1}
\mu^{\pm} = -4-\frac{3}{16\alpha}\pm\frac{1}{16\alpha}\sqrt{9+192\alpha}.
\end{equation}
The solution $\mathcal{V}^+$ is stable
for $\alpha > -9/192$. The solution $\mathcal{V}^-$ is a stable point
when $\alpha>0$ and unstable
for $-9/192 < \alpha < 0$.

\section{Non-vacuum solutions}

In this Section, we study the background dynamics of
models based on GB theories of the
type (\ref{model}) in the presence of a perfect fluid.
The vacuum solutions ${\cal{V}}^{\pm}$
remain as equilibrium points
of the autonomous system (\ref{autn1})-(\ref{autn1a}):
\begin{equation}
\label{vacV1}
(y_1,y_2) = \left(-\frac{1}{2}\pm\frac{1}{6}\sqrt{9+192\alpha}
,~~~\frac{3}{2}\mp\frac{1}{6}\sqrt{9+192\alpha}  \right).
\end{equation}
In addition, there exist two scaling solutions, where 
$\Omega_m$ and $\Omega_{\G}$ are constants: 
\begin{eqnarray}
\label{fps1}
(y_1, y_2)&=&\left(\pm\frac{2\sqrt{-3\alpha(1+3w_m)}}{3}
,~~~ \pm\frac{12\alpha(1+w_m)}{\sqrt{-3\alpha(1+3w_m)}}\right),
\\
\label{fps2}
\Omega_{m}&=&1 \mp\frac{2\sqrt{-3\alpha(1+3w_m)}}{3}
\mp \frac{12\alpha(1+w_m)}{\sqrt{-3\alpha(1+3w_m)}}, \\
\Omega_{\G}&=& \pm\frac{2\sqrt{-3\alpha(1+3w_m)}}{3}
\pm\frac{12\alpha(1+w_m)}{\sqrt{-3\alpha(1+3w_m)}}
\end{eqnarray}
and $w_{eff}=w_m$. 
We denote these solutions by  $\mathcal{S}^\pm$.

\begin{figure}[t]
\includegraphics[width=3.2in,height=2.3in]{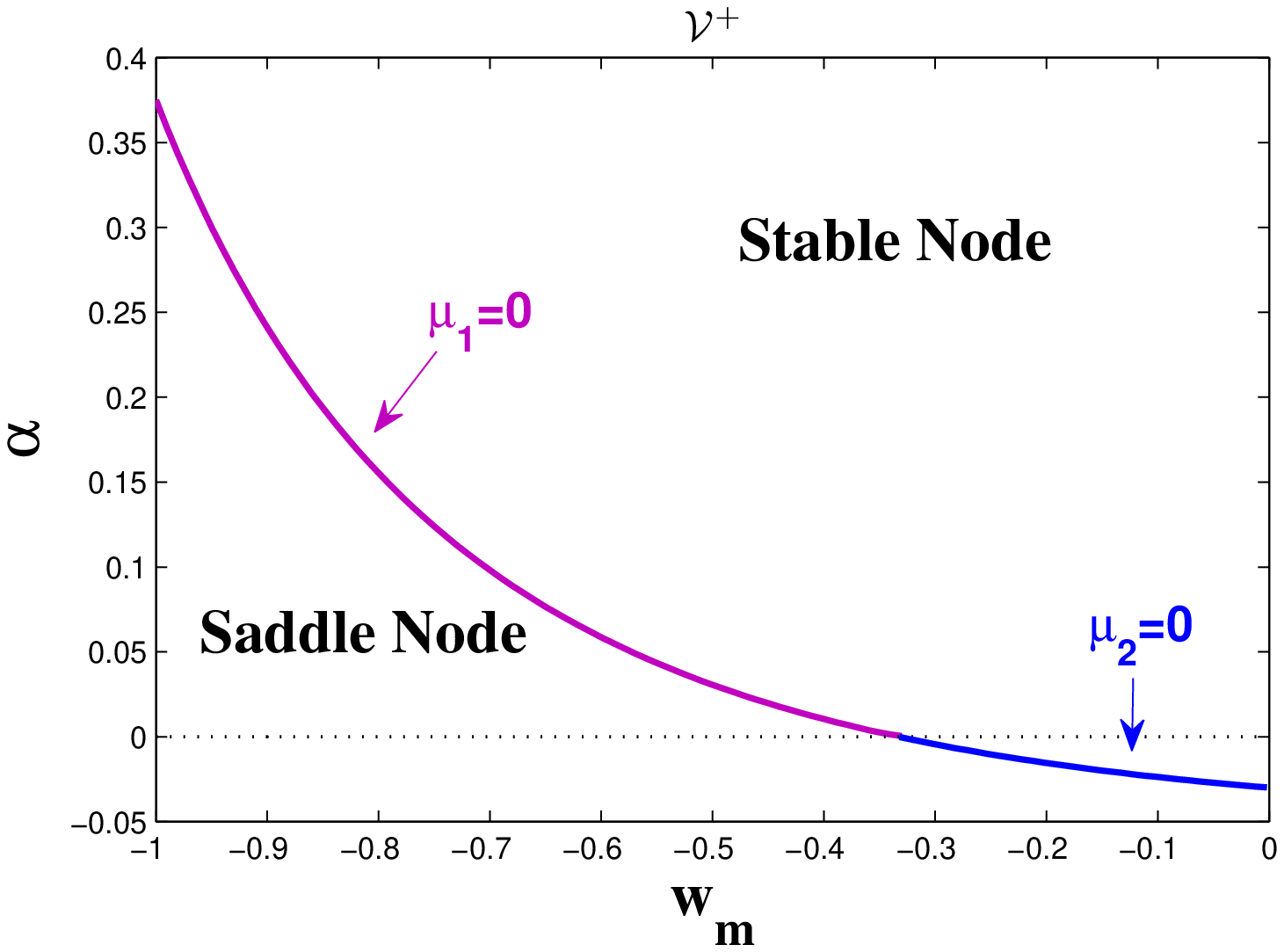}\includegraphics[width=3.2in,height=2.3in]{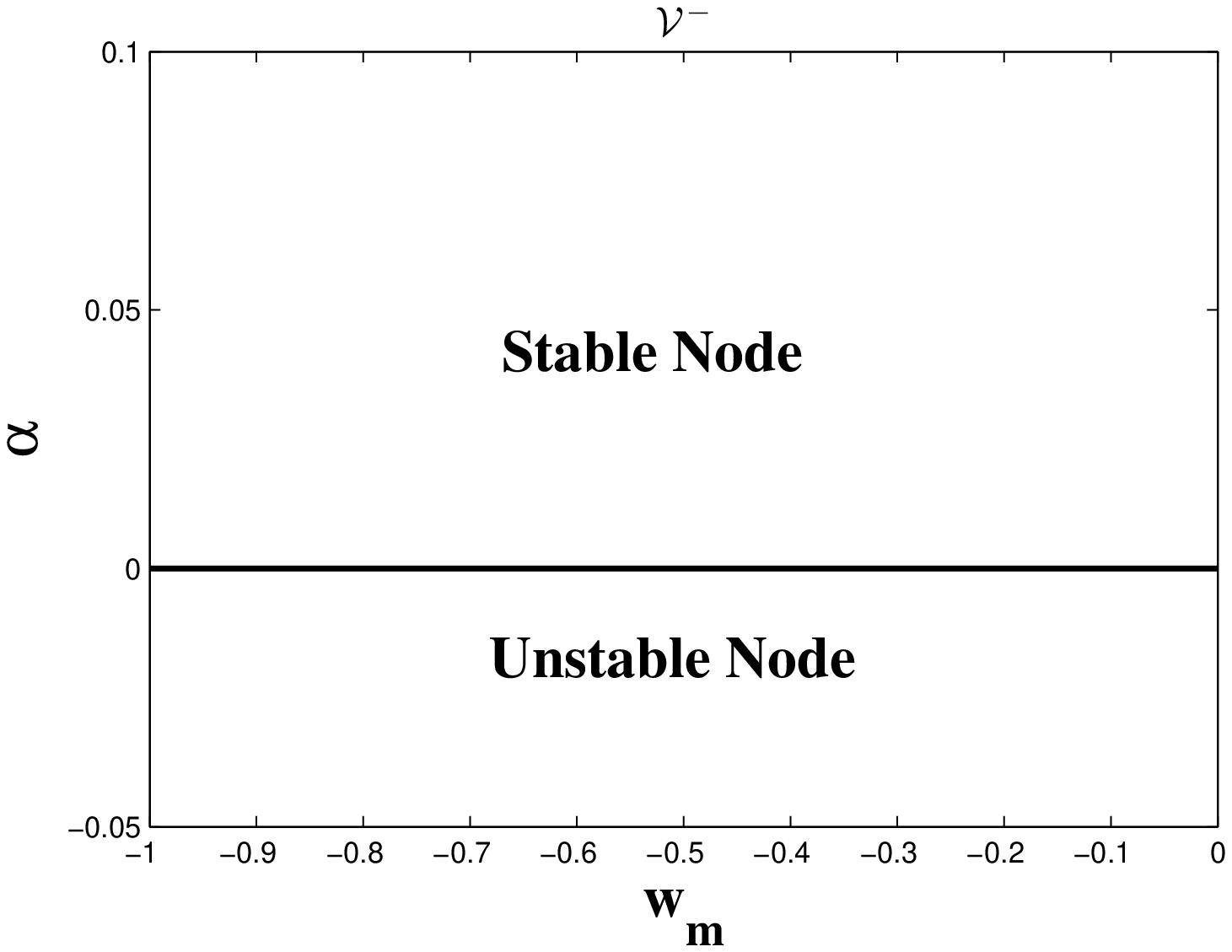}
\caption{\label{fig2}
Illustrating the nature of the equilibrium points ${\cal{V}}^{+}$
(left-hand panel) and ${\cal{V}}^{-}$
(right-hand panel) in the parameter space spanned by
$( w_m ,\alpha )$. Both fixed points
are real if $\alpha\geq-9/192$.  On the boundary (denoted by the solid line)
that distinguishes the
stability of the fixed point ${\cal{V}}^+$,
one of the eigenvalues $\mu_{1,2}^+$ vanishes. This is indicated
in the figure by a change in colour.
The dotted line in the left-hand panel represents the
invariant sub-manifold $y_1=0$.
In the case of the point $\mathcal{V}^-$, neither of the eigenvalues
vanishes in any region of the $(w_m , \alpha)$ plane.}
\end{figure}

The eigenvalues associated with the equilibrium points
${\cal{V}}^{\pm}$ are given by
\begin{eqnarray}
\label{ev1}
\mu_1^{\pm} &=& -\frac{1}{32\alpha} \left[ 48\alpha(3+w_m)+9 \mp 3\sqrt{9+192\alpha} \right]
+ \lambda_1^{\pm}  \\
\mu_2^{\pm} &=& -\frac{1}{32\alpha}\left[ 48\alpha(3+w_m)+9 \mp 3\sqrt{9+192\alpha} \right]
- \lambda_1^{\pm}
 \\
\lambda_1^{\pm}  &\equiv& \frac{1}{32\alpha} \left[ 256\alpha^2(1+3w_m)^2+288\alpha(1+w_m)
+18\mp32\alpha(1+3w_m)\sqrt{9+192\alpha} \mp 6\sqrt{9+192\alpha} \right]^{1/2}.
\end{eqnarray}
The stability of these vacuum solutions
is altered when a matter source
is introduced into the system and depends on both
the GB parameter, $\alpha$, and the perfect fluid equation of state, $w_m$.
This dependency is illustrated in Fig.~\ref{fig2}.
The solid lines represent the regions where
the nature of the equilibrium points changes as the parameter values are
altered. The stability of $\mathcal{V}^-$ is determined by the sign of the
GB parameter, $\alpha$. On the boundary distinguishing
the nature of the fixed point
$\mathcal{V}^+$, one of the eigenvalues $\mu_{1,2}^{+}$ vanishes.
To analyse the stability of the equilibrium point for these particular
choices of parameter values would require a second-order analysis,
which is beyond the scope of the present work.

\begin{figure}[t]
\includegraphics[width=3.2in,height=2.3in]{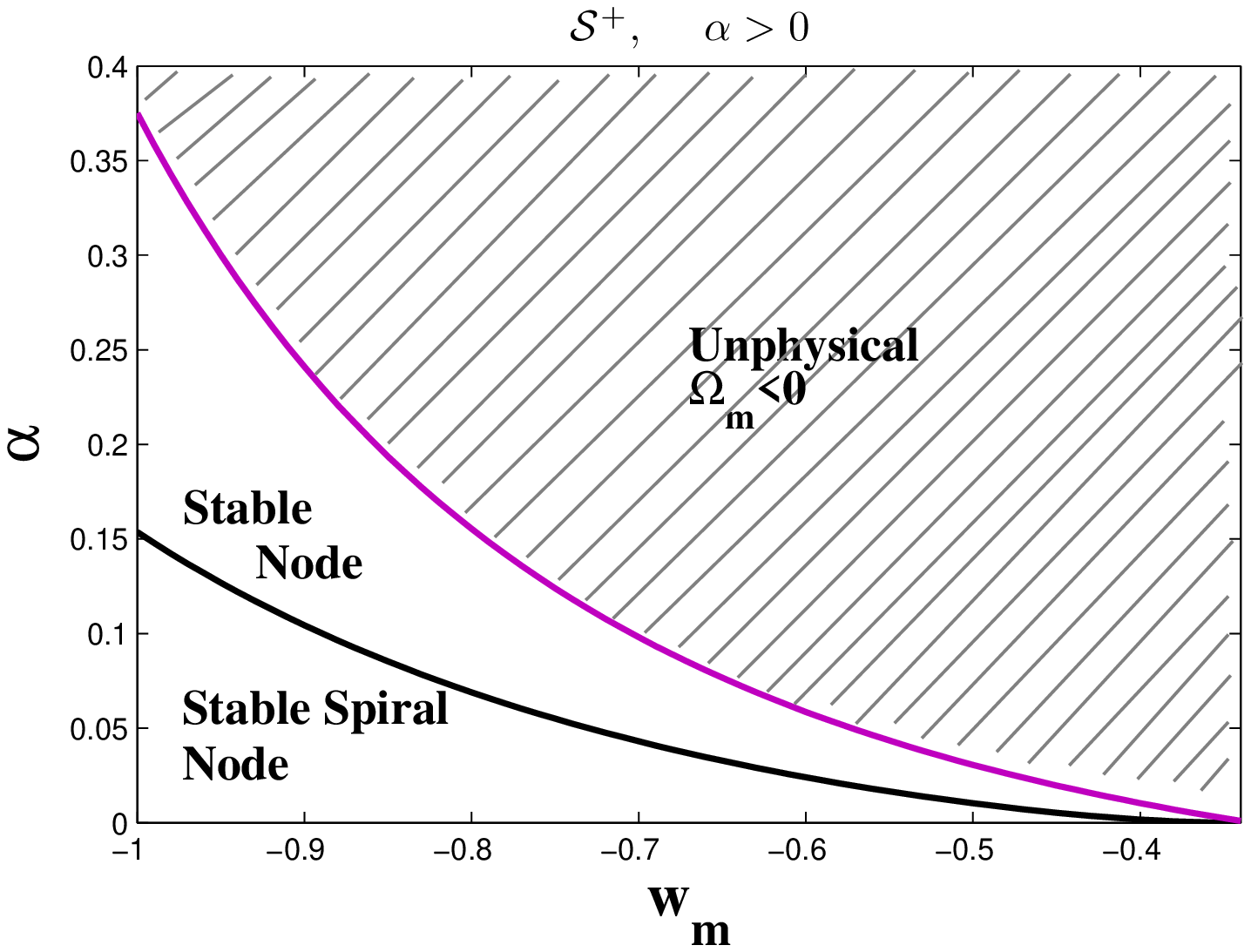}\includegraphics[width=3.2in,height=2.3in]{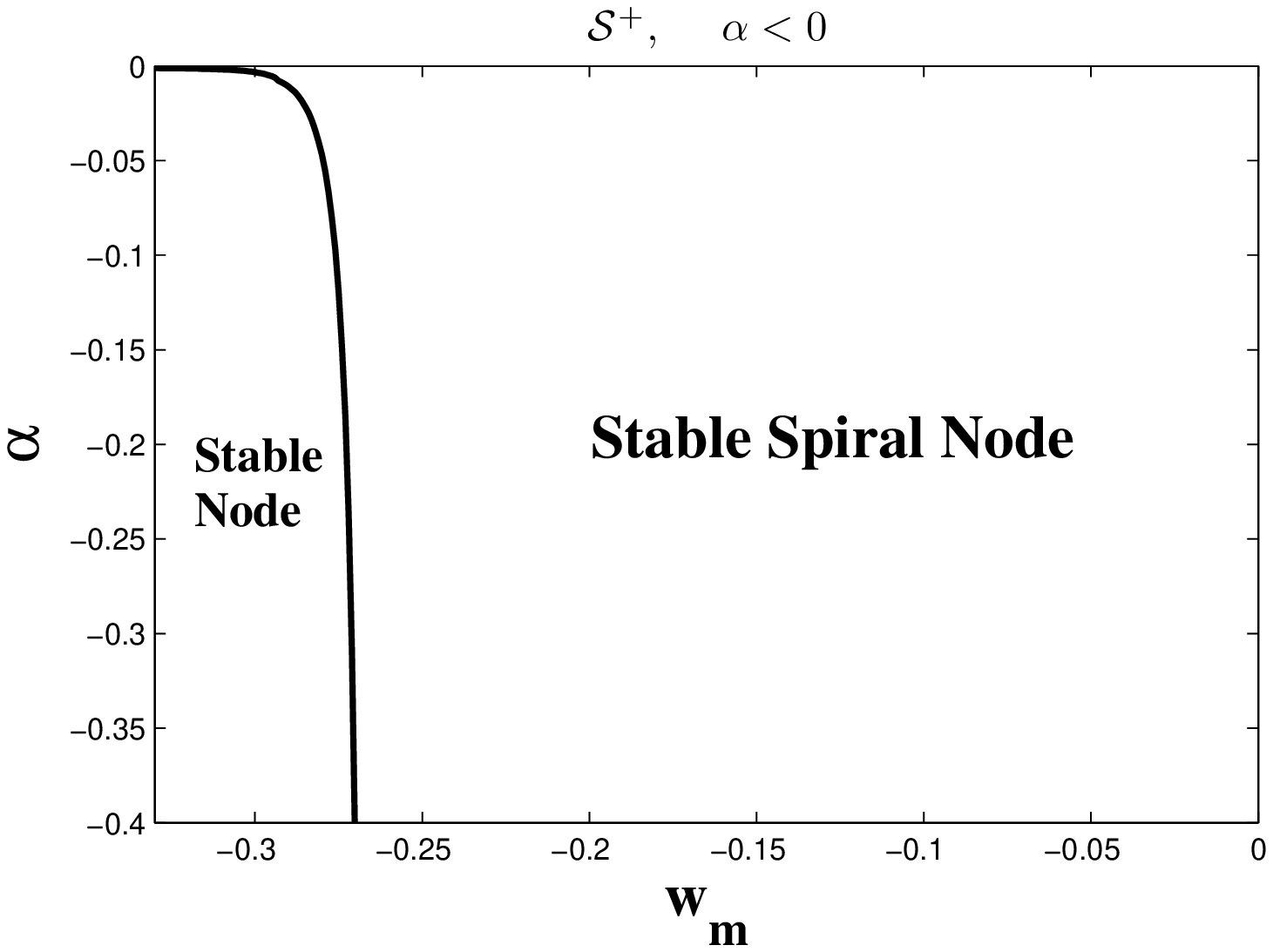}
\includegraphics[width=3.2in,height=2.3in]{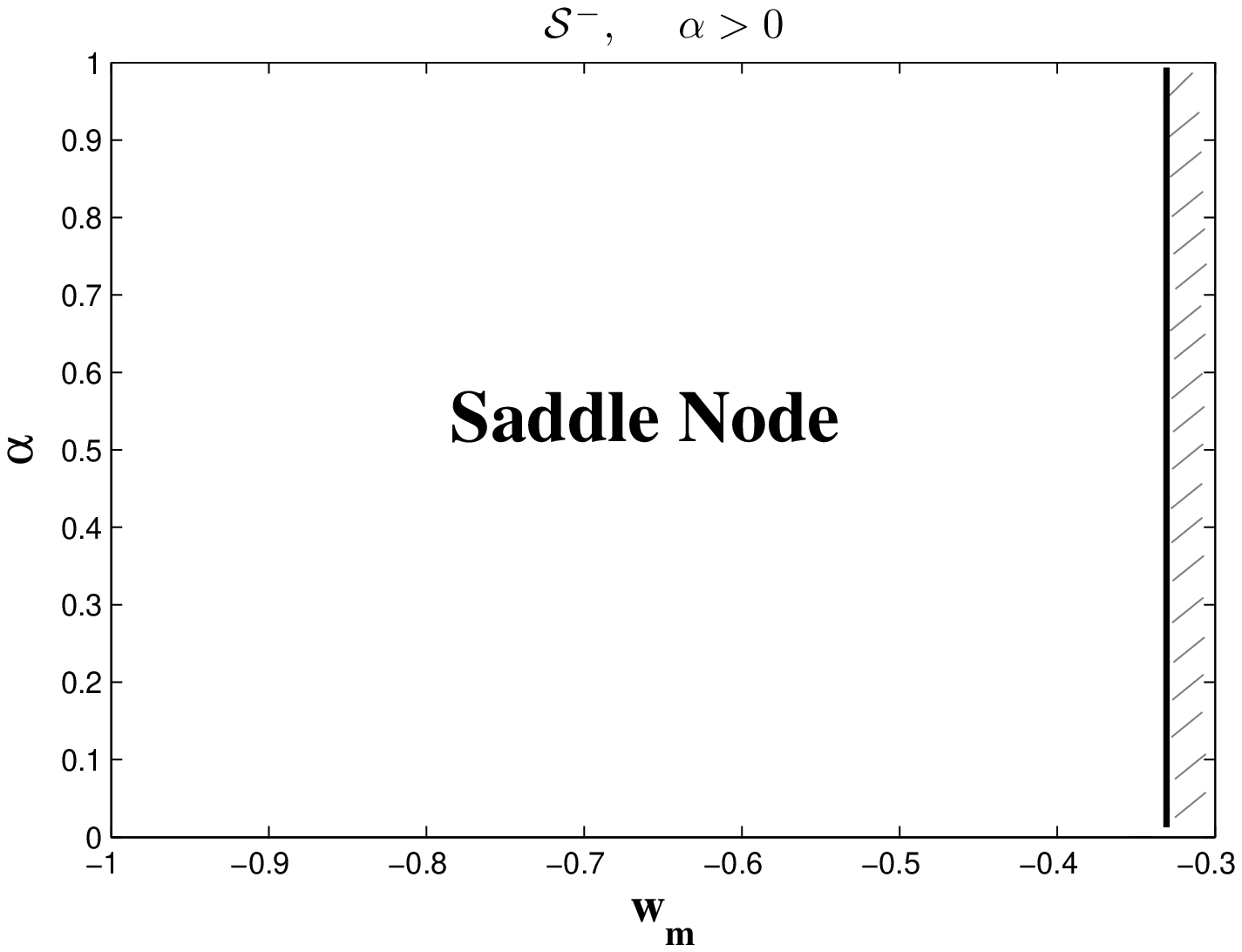}\includegraphics[width=3.2in,height=2.3in]{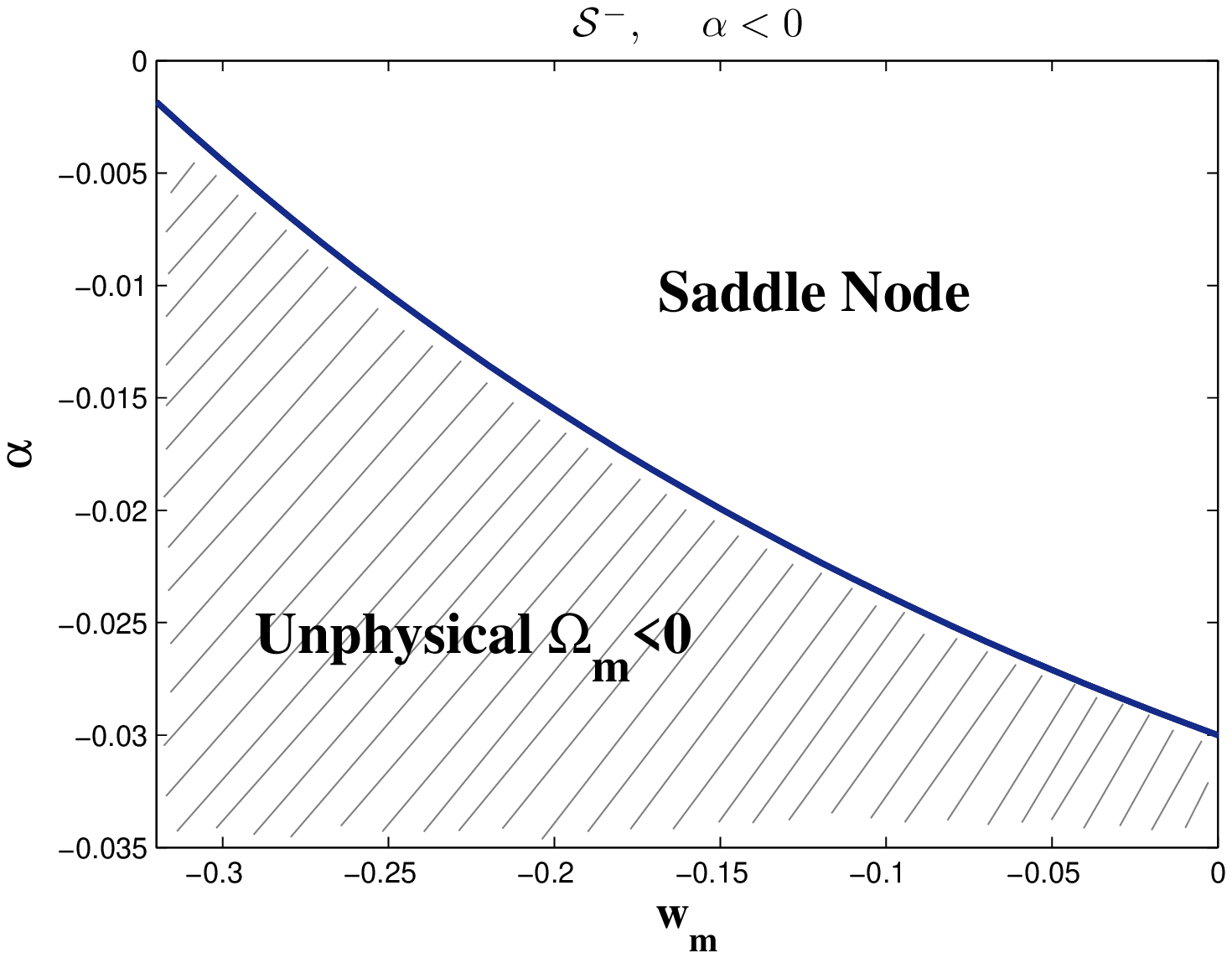}
\caption{\label{fig3}
Illustrating the stability of the scaling equilibrium points
${\cal{S}}^{\pm}$ in the parameter space spanned by $(w_m ,\alpha)$. The
region of parameter space is restricted by the requirement that the
equilibrium points are real, $\alpha(1+3w_m) \leq 0$, and also
correspond to physically realistic solutions where $\Omega_m \ge 0$.
The shaded areas depict the regions of parameter space where
the solutions are unphysical. These restrictions
imply that the analysis can be separated into
regions where $\alpha >0$ (left-hand panels) and $\alpha <0$ (right-hand panels).
The regions of parameter space
where the fixed points correspond to either a saddle point
or a stable/spiral node are identified. On the line $\Omega_m=0$,
the eigenvalue $\mu_1^+=0$ (for the scaling point ${\cal{S}}^+$)
 when $\alpha>0$. Conversely, $\mu_2^- =0$ (for the scaling point ${\cal{S}}^-$)
when $\Omega_m =0$ and $\alpha <0$.
}
\end{figure}

The eigenvalues associated with the scaling equilibrium points
${\cal{S}}^{\pm}$ are given by
\begin{eqnarray}
\label{es1}
\mu_1^{\pm} &=& \frac{3}{4}(w_m-1) + \tau_1^{\pm}  \\
\mu_2^{\pm} &=& \frac{3}{4}(w_m-1) - \tau_1^{\pm}  \\
\tau_1^{\pm} &\equiv& \frac{1}{4\alpha} \left[ \pm8\alpha(1+3w_m)\sqrt{-3\alpha(1+3w_m)}
-\alpha^2(135w_m^2+306w_m+71) \right]^{1/2}.
\end{eqnarray}
The stability of these fixed points is illustrated in Fig.~\ref{fig3}.
The points are real in the region of parameter space,
$\alpha(1+3w_m)\leq0$. Furthermore, they are only physically
meaningful if $\Omega_m = 1-y_1-y_2 \ge 0$. This results in a
further restriction in the $(w_m , \alpha)$ plane after substitution
of Eq. (\ref{fps2}).

The top two panels of Fig.~\ref{fig3} correspond
to the scaling solution $\mathcal{S}^+$ where $y_1>0$ and the bottom two
panels correspond to $\mathcal{S}^-$ where $y_1<0$. The point ${\cal{S}}^+$
is either a stable node or a stable spiral. The
point ${\cal{S}}^-$ is always a saddle. On
the curve $\Omega_m=0$, one of the eigenvalues of
$\mathcal{S}^\pm$ vanishes.

\begin{figure}[t]
\includegraphics[width=3.2in,height=2.3in]{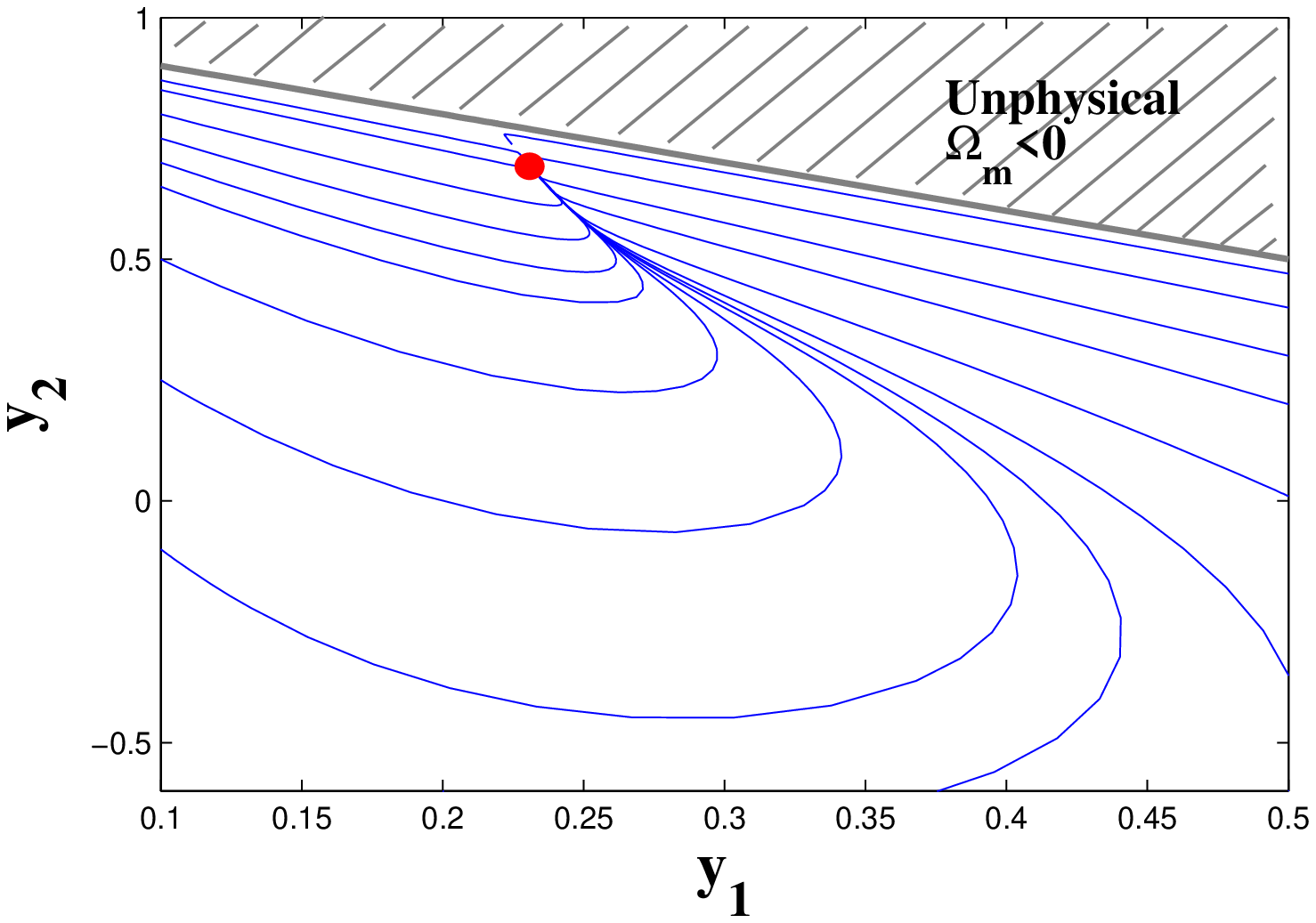}\includegraphics[width=3.2in,height=2.3in]{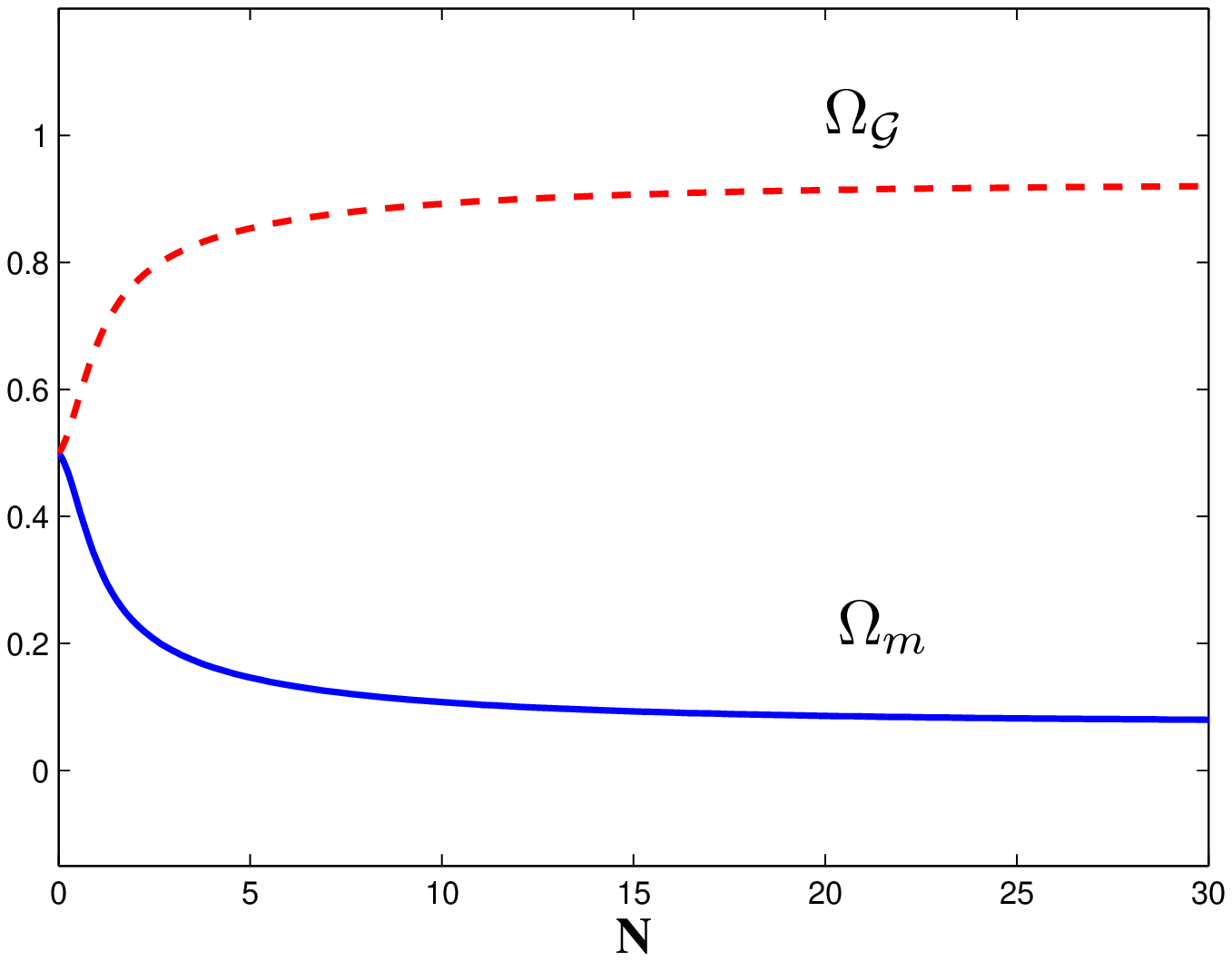}
\caption{\label{fig4}
Illustrating the dynamics of the model (\ref{model})
for the particular case where $(\alpha, w_m)=(0.05, -0.6)$.
The left-hand panel depicts the phase space, where the straight
line $y_1=1-y_2$ corresponds to the vacuum solution $\Omega_m=0$.
The red dot represents the scaling fixed point
$\mathcal{S}^+$.  For the range of
initial conditions chosen, all non-vacuum, physically acceptable solutions
are attracted to $\mathcal{S}^+$.
The right-hand panel depicts the evolution
of the fractional energy densities of the perfect
fluid, $\Omega_m$, and the GB contribution,
$\Omega_{\G}$, for the initial conditions $\Omega_m=\Omega_{\G}=0.5$.
It is seen that the fractional densities asymptote
to constant values at late times, thus
indicating that the solution is scaling.
}
\end{figure}

\begin{figure}[t]
\includegraphics[width=3.2in,height=2.3in]{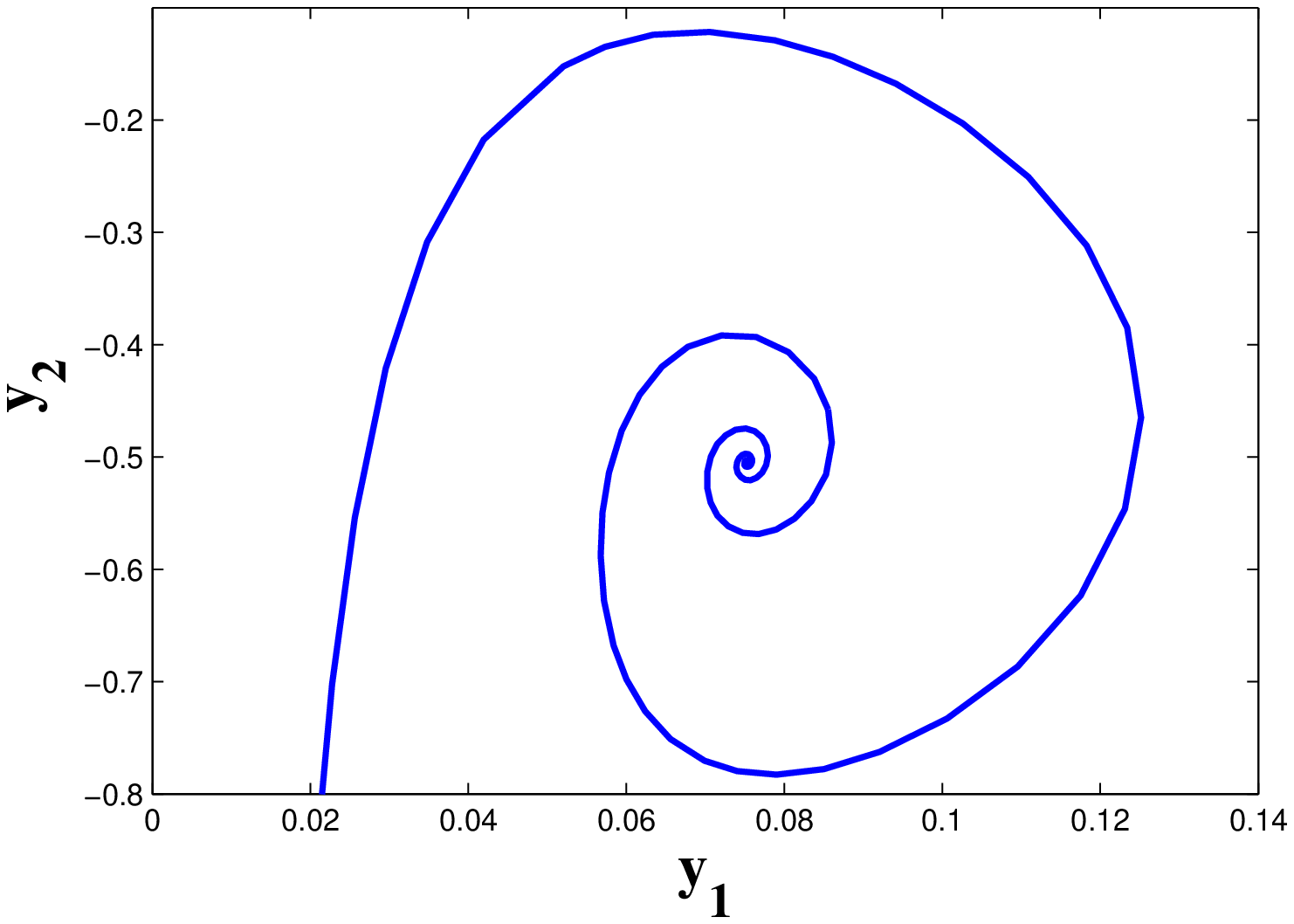}\includegraphics[width=3.2in,height=2.3in]{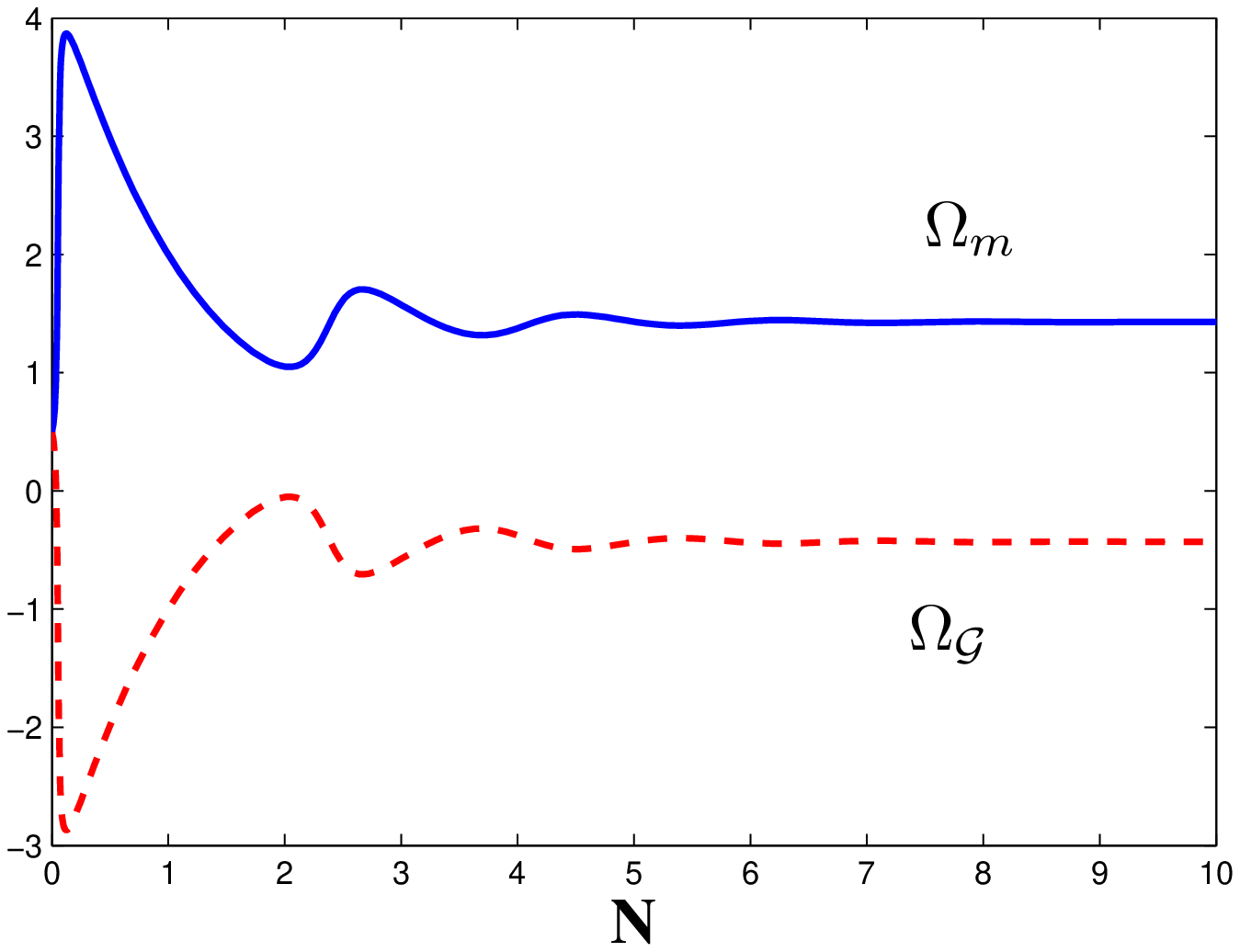}
\caption{\label{fig5}
Illustrating the dynamics of the model (\ref{model})
for the particular case where $(\alpha, w_m)=(-0.005, -0.05)$.
The left-hand panel depicts the phase space for this scenario,
whereas the right-hand panel depicts the evolution of the
fractional energy densities $\Omega_{m}$ and $\Omega_{\G}$.
The initial conditions were chosen such that $\Omega_m=\Omega_{\G}=0.5$.
At late times, the fractional energy densities of the fluid and
GB contribution tend to constant values.}
\end{figure}

To illustrate the scaling dynamics,
let us consider the specific case where $(\alpha, w_m)=(0.05, -0.6)$.
At this location in parameter space, there exist two equilibrium
points\footnote{Note that the point ${\cal{V}}^-$ also exists
but this occurs in the region $y_1 <0$. Stable scaling solutions arise
only for $y_1>0$ and, since $y_1 =0$ is a separatrix, a trajectory beginning
in the region $y_1<0$ will not be able to reach $\cal{S}^+$. We therefore
choose the initial conditions in Fig. \ref{fig4} such that $y_1 >0$.
This is equivalent to choosing the negative sign in Eq.~(23).}:
the saddle point $\mathcal{V}^+$ and the stable
node $\mathcal{S}^+$. The basin of attraction
for ${\cal{S}}^+$ is shown in Fig. \ref{fig4}.
As a second example, we consider the case $(\alpha, w_m)=(-0.005, -0.05)$,
where there exist four equilibrium points:
an unstable vacuum solution
$\mathcal{V}^-$, a saddle point $\mathcal{S}^-$,
a stable $\mathcal{V}^+$ and a stable spiral $\mathcal{S}^+$.
The spiral nature of the point $\mathcal{S}^+$ is illustrated
in the phase portrait of Fig.~\ref{fig5}, where the initial
conditions were specified to be $\Omega_m=\Omega_{\G}=0.5$.

\section{Conclusion}

In this paper we have investigated the existence and stability 
of cosmological power-law scaling solutions sourced by  
a barotropic fluid when an appropriate function 
of the Gauss-Bonnett topological invariant is introduced 
into the Einstein-Hilbert action. It was found that 
the general class of such theories that 
admit power-law solutions is given by Eq. (\ref{model}), 
i.e.,  $f(\G)= \pm 2 \sqrt{\alpha \G}$ for some constant 
coefficient, $\alpha$. By exploiting an equivalence between generalized
Gauss-Bonnet gravitational theories and a corresponding higher-order, 
scalar-tensor theory, it was further shown that the Friedmann equations 
for this class of model can be written in the form of a two-dimensional 
dynamical system. The stability of the equilibrium points for 
both vacuum and non-vacuum models was established. 
In the former case, the GB parameter, $\alpha$, determines 
the effective equation of state parameter. For non-vacuum 
solutions, the nature of the critical points depends on both  
$\alpha$ and the fluid equation of state parameter, $w_m$. The regions of
parameter space $(\alpha, w_m)$ that admit stable
non-vacuum scaling solutions were identified. 

The models we have investigated do not admit a transition 
from a decelerating to an accelerating phase of cosmic expansion. However, 
our aim in this paper has been to focus on power-law solutions rather than 
develop a phenomenological model of generalized Gauss-Bonnet gravity as a 
candidate for dark energy. Power-law solutions are of interest since they 
can be regarded as approximations to more realistic models. In particular, 
phenomenological models could be constructed where the parameter $\alpha$ 
is given by some function of $\G$ 
(or equivalently the scalar field $\phi$),  
such that $\alpha$ is slowly varying for much of the 
history of the universe, but at some epoch undergoes a change in sign.
In principle, this could cause the universe to enter a phase of accelerated 
expansion. It would be interesting to develop specific models
of this type, along the lines outlined in Ref.~\cite{DeFelice}.

\section*{Acknowledgements}
We thank Shinji Tsujikawa for useful discussions
and Baojiu Li for helpful comments. K.~U. is supported
by the Science and Technology Facilities Council (STFC).


\end{document}